\documentclass{article}

\usepackage{arxiv}

\usepackage[utf8]{inputenc} % allow utf-8 input
\usepackage[T1]{fontenc}    % use 8-bit T1 fonts
\usepackage{hyperref}       % hyperlinks
\usepackage{url}            % simple URL typesetting
\usepackage{booktabs}       % professional-quality tables
\usepackage{amsfonts}       % blackboard math symbols
\usepackage{nicefrac}       % compact symbols for 1/2, etc.
\usepackage{microtype}      % microtypography
\usepackage{lipsum}

\usepackage{graphics} % for pdf, bitmapped graphics files
\usepackage{epsfig} % for postscript graphics files
\usepackage{times} % assumes new font selection scheme installed
\usepackage{amsmath} % assumes amsmath package installed
\usepackage{amssymb}  % assumes amsmath package installed
\usepackage{booktabs}
\usepackage[dvipsnames]{xcolor}
\usepackage{siunitx}
\usepackage{amsfonts}

\newtheorem{problem}{Problem}
\newtheorem{theorem}{Theorem}
\newtheorem{remark}{Remark}
\newenvironment{proof}{\paragraph{Proof:}}{\hfill$\square$}
\newtheorem{assumption}{Assumption}
\newtheorem{lemma}{Lemma}
\DeclareMathAlphabet{\mathbbold}{U}{bbold}{m}{n}
\usepackage{enumitem}   
\usepackage{algorithm}
\usepackage[noend]{algpseudocode}
\algrenewcommand\algorithmicrequire{\textbf{Input:}}
\algrenewcommand\algorithmicensure{\textbf{Output:}}

\usepackage{subcaption}
\usepackage{graphicx}
\usepackage{textcomp}
\usepackage{xcolor}
\usepackage{svg}
\usepackage{bbm}

\algdef{SE}[SUBALG]{Indent}{EndIndent}{}{\algorithmicend\ }%
\algtext*{Indent}
\algtext*{EndIndent}

\title{\LARGE \bf
Co-design of Output-Feedback Neural Network Control Barrier Functions and Observers for Stochastic Jump-Diffusion Systems
\thanks{This work was supported by ANRF ARG grant and Siemens fellowship.}
}

\author{
 Vaishnavi Jagabathula\\
  Centre for Cyber-Physical Systems\\
  IISc, Bengaluru, India\\
  \texttt{vaishnavij@iisc.ac.in} \\
   \And
 Ahan Basu \\
  Centre for Cyber-Physical Systems\\
  IISc, Bengaluru, India\\
  \texttt{ahanbasu@iisc.ac.in} \\
  \And
 Pushpak Jagtap \\
  Centre for Cyber-Physical Systems\\
  IISc, Bengaluru, India\\
  \texttt{pushpak@iisc.ac.in} \\
  }

\begin{document}
\maketitle
\thispagestyle{empty}
\pagestyle{empty}
\begin{abstract}
In this paper, we present a framework for designing controllers for partially observed continuous-time stochastic systems. We use the notion of control barrier functions (CBF) and provide an upper bound to the probability of state trajectories violating the system's safety when starting in an initial region. Our framework co-designs a neural network-based controller, an output-feedback observer, and a control barrier function for the given system. Although CBF requires full state information, we formulate CBF conditions over an augmented state space of actual and observer states. We train the neural networks simultaneously by defining appropriate loss functions, and finally demonstrate the effectiveness through two case studies: a linear DC motor and a nonlinear system. 
    % Control Barrier Functions (CBFs) provide a powerful framework for ensuring safety in dynamical systems. However, their application typically relies on full state information, which is often violated in real-world scenarios due to the availability of partial state information. In this work, we propose a neural network-based framework for the co-design of a safety controller, observer, and CBF for partially observed continuous-time systems. By formulating barrier conditions over an augmented state space, our approach ensures safety without requiring bounded estimation errors or handcrafted barrier functions. All components are jointly trained by formulating appropriate loss functions, and we introduce a validity condition to provide formal safety guarantees beyond the training data. Finally, we demonstrate the effectiveness of the proposed approach through two case studies: a DC motor and a pendulum model.
\end{abstract}
%\textbf{\textit{Keywords—} Partially observed systems, control barrier functions, controller synthesis, output-feedback observer, neural network learning, formal guarantees.}
%\begin{IEEEkeywords}
%Partially observed systems, control barrier functions, controller synthesis, output-feedback observer, neural network learning, formal guarantees.
%\end{IEEEkeywords}
\section{Introduction}
The safety of dynamical systems deployed in the real world remains a crucial research problem. Incorporating safety criteria into the controller design framework is a common approach. Researchers have proposed barrier certificates for safety validation of continuous and hybrid systems in \cite{PRAJNA2006117}. This barrier certificate framework was subsequently extended to worst-case and stochastic safety verification in \cite{prajna_stoch}, providing sufficient conditions to establish safety and to bound the probability of state trajectories reaching unsafe regions. Building on this probabilistic foundation, Control Barrier Certificates (CBC) have been studied as an approach to safety-critical controller synthesis for stochastic systems \cite{cbf_stochastic}, enabling the computation of upper bounds on the probability of safety violation over a finite time horizon. In parallel, an alternative CBF framework \cite{ames2019control} frames safety in the context of enforcing set invariance for deterministic continuous-time systems, enabling controller synthesis via quadratic programming. However, this approach requires prior knowledge of a CBF and may be limited in the presence of control input constraints. Returning to the stochastic setting, CBCs have been used to synthesize control strategies for stochastic control-affine systems \cite{cbf_stochastic, wang2021safety, 10559418} and non-affine systems \cite{cbf_stochastic, vahs2024non}.

Despite their wide applicability, conventional CBF-based methods rely on full state information. In practical settings, such assumptions are prone to being violated due to inherently unobserved states, and as a consequence, applying CBFs directly to partially observed systems can degrade performance and compromise safety constraints. 

Safety-critical control of nonlinear stochastic systems under partially observed states remains a fundamental problem. In \cite{zhang2025conformal}, the authors proposed a sampled-data, measurement-robust risk-aware CBF framework for stochastic systems with noisy state measurements. By quantifying the estimation error using conformal prediction, the proposed approach establishes an upper bound on the probability of state trajectories entering an unsafe region over a finite time duration. However, the framework assumes that a suitable control barrier function is available a priori, requiring manual construction for the CBF. An alternative approach includes the formulation of barrier conditions on an augmented system of actual and estimated dynamics \cite{jump_diff_sys,jahanshahi2020synthesis}. However, such methods face key limitations: the assumption of the availability of state observers, the estimation error bounds are often difficult to guarantee in practice, and accurate observers for highly nonlinear systems are rarely feasible. Recently, the construction of neural CBFs has emerged as a promising alternative in synthesizing the safety controller, eliminating the need for manually designed barrier functions. Neural network-based \cite{anand2023formally,tayal2024learning,11755823} and data-driven approaches \cite{Data_CBF_Zamani, jahanshahi2023synthesis,jagtap2020control} to construct CBFs for unknown or partially observed systems have gained significant attention, but still require either the full-state information of the system or bounded estimation error. {Our previous work addressed the synthesis of neural control barrier functions and controllers for a continuous-time system \cite{NN_observer_cts_time}. But for stochastic jump-diffusion systems, it is not straightforward to ensure safety over the complete time horizon. The controller for stochastic jump-diffusion systems can only provide a safety guarantee for a finite time horizon. As a result, designing such controllers is not straightforward compared to the previous work.}

In this work, we propose a neural network-based framework for the co-synthesis of a safety controller with an observer and CBF for partially observed continuous-time stochastic systems and provide an upper bound on the probability of starting in an initial region and entering an unsafe region within a time interval. Leveraging the universal approximation capability of neural networks, the core idea is to formulate CBF conditions over an augmented state space that satisfy the safety of the latent system. In our setting, we adopt a co-design strategy such that the controller explicitly depends on the observer’s state estimates, enabling a more cohesive and effective synthesis. In addition to this, the proposed CBF formulation is designed to be independent of the observer’s estimation error, thereby eliminating the need for an exact or explicitly accurate observer. To implement this framework, we present the barrier, the controller, and the observer as neural networks and train them simultaneously to ensure the safety guarantee of the actual system.

\section{Preliminaries and problem statement}
\textit{Notations.}
The symbols $\mathbb{R}, \mathbb{R}_{\geq 0}$ denote the set of real and non-negative real numbers. A column vector with $n$ rows of real number entries $x_1,..., x_n$ is denoted as $x=[x_1,...,x_n]^\top$. The $n$-dimensional vector space is represented by $\mathbb{R}^n$. The standard basis vector is denoted by $\mathbf{e}_i$, where the $i$-th element is 1, and the rest are zero. The vector space of real matrices with $n$ rows and $m$ columns is represented by $\mathbb{R}^{n\times m}$. Given a matrix $A\in \mathbb{R}^{m\times n}$, we denote the trace of $A$ as Tr$(A)$. The first-order partial differentiation of a function $f:X\times \hat{X}\rightarrow \mathbb{R}$ with respect to the variable $x\in X$ and $\hat{x} \in X$ are denoted by $\partial_x f$ and $\partial_{\hat{x}}f$ respectively. The second-order partial differentiation with respect to $x$ is denoted by $\partial_{xx}f$. %The Hessian of the function $f$ is denoted by $\nabla^2f$. 
The Euclidean norm is represented using $\lVert \cdot \rVert$. The indicator function is defined as $\mathbbold{1}_{x\in X} =1$, if $x\in X$, $0$ otherwise. A function $f$ is Lipschitz continuous with Lipschitz constant $L$ if $\lVert f(x_1)-f(x_2)\rVert\leq L\lVert x_1-x_2\rVert$. The set of natural numbers from $1$ to $N$ is denoted by $[1;N]$.

\subsection{System Description and Problem Formulation}
In this work, we consider a continuous-time stochastic jump-diffusion system (ct-SJDS). Let ($\Omega,\mathcal{F},\mathbb{P}$) be the probability space, where $\Omega$ is a sample space, $\mathcal{F}$ is the filtration satisfying the conditions of completeness and right continuity, and $\mathbb{P}$ is the associated probability measure. For $k\in \{1,2\}$, let us denote $\overline{r}_k$-dimensional $\mathcal{F}$-Brownian motions as $(W_{ks})_{s\geq 0}$ and $\overline{q}_k$-dimensional $\mathcal{F}$-Poisson processes as $(P_{ks})_{s\geq 0}$ and they are independent of each other. The Poisson processes $P^i_{ks}$ for $i\in\{1,...,\overline{q}_k\}, k=\{1,2\}$ have the rates $\lambda_{ki}$.
Under these considerations, we consider the partially observed ct-SJDS described by the following stochastic differential equations:
\begin{align}
    S:=\begin{cases}
    \text{d}{x}= f(x,u)\text{d}t +g_1(x)\text{d}W_{1t}+r_1(x)\text{d}P_{1t}, \\
    \text{d}y = h(x)\text{d}t,%+\textcolor{blue}{g_2}\text{d}W_{2t}+r_2\text{d}P_{2t}, 
    \end{cases}\label{dyn}
\end{align}
with state $x(t)\in X\subset \mathbb{R}^n$ under an input $u(t)\in U\subset \mathbb{R}^m$ at time $t\geq 0$. The output at time $t\in\mathbb{R}_{\geq 0}$ is denoted as $y(t)\in Y\subset \mathbb{R}^p$\ $\mathbb{P}$-almost surely ($\mathbb{P}$-a.s). The functions $f:X\times U\rightarrow\mathbb{R}^n, g_1:X\rightarrow\mathbb{R}^{n\times \overline{r}_1}, r_1:X\rightarrow\mathbb{R}^{n\times \overline{q}_1},$ and $h:X\rightarrow\mathbb{R}^p$ are assumed to be Lipschitz continuous functions to ensure the existence and uniqueness of the solution \cite{oksendal2019applied}. We denote the value of the solution process of $S$ at time $t\in \mathbb{R}_{\geq 0}$, starting at initial state $x_0, \mathbb{P}$-a.s. under the control signal $v$ as $x_{x_0,v}(t)$. Here, we consider partially observed systems in which the full-state measurements are unavailable (i.e., $p<n$).

\begin{problem}
    Given a partially observed ct-SJDS $S$ as in \eqref{dyn}, an initial region $X_0\subset X$, and an unsafe region $X_u\subset X$, such that $X_0\cap X_u=\emptyset$, our objective is to design an output-feedback controller for the system that guarantees a potentially tight lower bound on the probability of the system trajectory starting from $x_0\in X_0$ $\mathbb{P}$-a.s. does not enter the unsafe set $X_u$ under the control policy $v$ for the time duration $[0,T)$, i.e., $x_{x_0,v}(t)\notin X_u, \forall t\in [0,T)$.
\end{problem}

Since we are dealing with a partially observed system, in order to design a suitable controller, we need to estimate the states using an observer. We assume that the observer designed for $S$ has the following dynamics:
% \begin{align}
%     \hat{S}:=\begin{cases}
%     \text{d}{\hat{x}} = \hat{f}(\hat{x},u, \text{d}y-\text{d}\hat{y})\text{d}t,\\
%     \text{d}\hat{y} = h(\hat{x})\text{d}t.
%     \end{cases}\label{dyn_obs}
% \end{align}
\begin{align}
    \hat{S}:=%\begin{cases}
    \text{d}{\hat{x}} = \hat{f}(\hat{x},u, h(x)-h(\hat x))\text{d}t,%\\
    % \text{d}\hat{y} = h(\hat{x})\text{d}t.
    % \end{cases}
    \label{dyn_obs}
\end{align}
Next, we define the augmented state $\text{d}\tilde{x} = [\text{d}x^\top, \text{d}\hat{x}^\top]^\top$, where $\text{d}x$ and $\text{d}\hat{x}$ are the states of $S$ and $\hat{S}$, respectively. The corresponding augmented system $\tilde{S}$ is defined as:
\begin{align}
    \begin{bmatrix}
        \text{d}{x}\\\text{d}{\hat{x}}
    \end{bmatrix}&=\begin{bmatrix}
        f(x, u)\\ \hat{f}(\hat{x},u, h(x)-h(\hat x))
    \end{bmatrix}\text{d}t+
    \begin{bmatrix}
         g_1(x)\\ 0_{n\times \overline{r}_1}
    \end{bmatrix}\text{d}{W_{1t}}+\begin{bmatrix}
        r_1({x})\\0_{n\times \overline{q}_1}
    \end{bmatrix}\text{d}P_{1t}.
\end{align}

Existing observer-based approaches rely on the exact structure of the observer, and designing observers for general nonlinear jump-diffusion systems becomes challenging and remains an open problem. To overcome this, we adopt a neural network-based observer whose parameters are trainable.
%\VJ{In [9], the observer is neural network based} 
\subsection{Control Barrier Function (CBF) for Augmented System}
We consider an approach that utilizes the notion of control barrier functions to develop a controller to keep the actual system \eqref{dyn} away from unsafe regions. In this section, we provide sufficient conditions using control barrier functions (CBFs) for the augmented system (inspired by the work in \cite{jump_diff_sys}), and provide an upper bound on the probability that the system trajectories starting from an initial state in the compact set $X_0\subseteq X$ reach the unsafe region $X_u\subseteq X$. Suppose, there exists a continuously differentiable function $B:X\times X\rightarrow \mathbb{R}_{\geq 0}$ and a continuous function $\textsl{g}:X\rightarrow U$, which satisfies the following conditions:
\begin{subequations}
\begin{align}
    \forall& (x, \hat{x})\in X_0\times X_0, B(x,\hat{x}) \leq \gamma_1, \label{cbf_1}\\
    \forall& (x, \hat{x})\in X_u\times X, B(x,\hat{x}) \geq \gamma_2 ,\label{cbf_2}\\
    \forall &(x, \hat{x})\in X\times X, \mathcal{D}B(x,\hat{x},\ {\textsl{g}(\hat x)})\leq c,\label{cbf_3}
\end{align}
\end{subequations}
where $\gamma_1\in[0,1], \gamma_2> \gamma_1$ and $c\geq 0$ are constants and $\mathcal{D}B$ is given as:
\begin{align}
    \mathcal{D}B &= \partial_x Bf(x,{\textsl{g}(\hat x)}) + \partial_{\hat{x}}B\hat{f}(\hat{x},{\textsl{g}(\hat x)}, h(x)-h(\hat x)) +\frac{1}{2}\text{Tr}(g_1(x)g_1(x)^\top \partial_{xx}B(x,\hat x)) \nonumber\\&+\sum_{i=1}^{\overline{q}_1} \lambda_{1i}(B(x+r_1(x)\mathbf{e}_i, \hat x)-B(x,\hat x)).%\nonumber\\&+\sum_{i=1}^{\overline{q}_2} \lambda_{2i}(B(x+r_2\mathbf{e}_i, \hat x)-B(x,\hat x))
    \label{db}
\end{align}

Now we state the main theorem of the section that provides an upper bound on the reachability probability for the trajectory of the system $S$.

\begin{theorem}\label{th:barrier}
    Consider the partially observed ct-SJDS $S$, as in \eqref{dyn}, its state observer $\hat{S}$ as in \eqref{dyn_obs}, and the augmented system $\tilde{S}$. The initial and unsafe sets are $X_0, X_u\subseteq X$, {such that $X_0\cap X_u=\emptyset$}. Suppose that a continuously differentiable function $B$ exists under the controller $\textsl{g}$ satisfying conditions \eqref{cbf_1}-\eqref{cbf_3}. Then, the probability that the solution process $x_{x_0,v}(t)$ of the system starting from any initial state $x_0\in X_0$ and reaching region $X_u$ under the control policy $v(t)={\textsl{g}(\hat x(t))}$ within the time duration $[0,T)\subset\mathbb{R}_{\geq 0}$ is upper bounded by $\frac{\gamma_1}{\gamma_2} +cT$. 
\end{theorem}
\begin{proof} 
We have the CBF conditions \eqref{cbf_1}-\eqref{cbf_3} and $X_u\times X\subseteq \{(x,\hat{x}\in X\times X \mid B(x,\hat x)\geq \gamma_2\}$. Using monotonicity property of probability, we have $\mathbb{P}\{x_{x_0,v}(t)\in X_u \land \hat{x}_{\hat{x}_0,v}(t)\in X, \exists t\in[0,T)\mid x_0, \hat{x}_0\}\leq \mathbb{P}\{\sup_{0\leq t\leq T}B(x_{x_0,v}(t), \hat{x}_{\hat{x}_0,v}(t))\geq 
\gamma_2 \mid x_0, \hat{x}_0\}$. Using Lemma 1 in \cite{kushner2006stochastic}, we obtain $\mathbb{P}\{\sup_{0\leq t\leq T}B(x_{x_0,v}(t), \hat{x}_{\hat{x}_0,v}(y))\geq 
\gamma_2 \mid x_0, \hat{x}_0\}\leq \frac{B(x_0, \hat{x}_0)}{\gamma_2}+cT\leq \frac{\gamma_1}{\gamma_2}+cT$. Now, $\mathbb{P}\{x_{x_0,v}(t)\in X_u \land \hat{x}_{\hat{x}_0,v}(t)\in X, \exists t\in[0,T)\mid x_0, \hat{x}_0\}=\mathbb{P}\{x_{x_0,v}(t)\in X_u, \exists t\in[0,T)\mid x_0, \hat{x}_0\} + \mathbb{P}\{\hat{x}_{\hat{x}_0,v}(t)\in X, \exists t\in[0,T)\mid x_0, \hat{x}_0\}- \mathbb{P}\{x_{x_0,v}(t)\in X_u \vee \hat{x}_{\hat{x}_0,v}(t)\in X, \exists t\in[0,T)\mid x_0, \hat{x}_0\}$. The probabilities $\mathbb{P}\{\hat{x}_{\hat{x}_0,v}(t)\in X, \exists t\in[0,T)\mid x_0, \hat{x}_0\}=1$ and $\mathbb{P}\{x_{x_0,v}(t)\in X_u \vee \hat{x}_{\hat{x}_0,v}(t)\in X, \exists t\in[0,T)\mid x_0, \hat{x}_0\}=1$, we get $\mathbb{P}\{x_{x_0,v}(t)\in X_u, \exists t\in[0,T)\mid x_0, \hat{x}_0\}=\mathbb{P}\{x_{x_0,v}(t)\in X_u \land \hat{x}_{\hat{x}_0,v}(t)\in X, \exists t\in[0,T)\mid x_0, \hat{x}_0\}\leq \frac{\gamma_1}{\gamma_2}+cT$. This concludes the proof.
\end{proof}

\section{Neural Network-based Observer, CBF and Controller with Formal Guarantees}
In this section, we present our proposed framework for synthesizing control barrier functions (CBF), an observer, and a state estimation-based controller for a partially observed ct-SJDS. We first reformulate the CBF conditions \eqref{cbf_1}-\eqref{cbf_3} as an optimization problem. We then address the challenges to solve the optimization problem and propose the CBF, controller, and observer as neural networks and derive the corresponding loss functions. Finally, we present the simultaneous training procedure for all three neural networks. 

Since Theorem \ref{th:barrier} establishes probability of state trajectories reaching the unsafe region is upper bounded by $\gamma_1/\gamma_2+cT$, we design our objective function to minimize this upper bound by decreasing $\gamma_1,c$, and increasing the gap between safe and unsafe sets denoted by $\gamma_2-\gamma_1$:
\begin{align}
     &\min_{\gamma_1, \gamma_2, c, B, \textsl{g}, \hat{f}} c-\beta(\gamma_2-\gamma_1)\label{op_prob}\\
     \text{s.t. }& \forall (x,\hat{x})\in X\times X, \max(q_i(x,\hat x))\leq 0,\nonumber\\
     &q_1(x,\hat x) = (B(x,\hat{x})-\gamma_1)\mathbbold{1}_{(x,\hat{x})\in X_0\times X_0}, \nonumber\\ 
     &q_2(x,\hat x)= (-B(x,\hat{x}) +\gamma_2)\mathbbold{1}_{(x,\hat{x})\in X_u\times X}, \nonumber\\
     &q_3(x,\hat x) = (\mathcal{D}B(x, \hat{x},\textsl{g}(\hat x))- c)\mathbbold{1}_{(x,\hat{x})\in X\times X}, \nonumber\\
     & 0\leq \gamma_1<1, \gamma_2> \gamma_1, c\geq 0, \nonumber
 \end{align} 
where $\beta>0$ is a user-defined parameter.
We provide the following lemma about the barrier, controller, and observer. However, such methods face key limitations: the assumption of the availability of state observers, the estimation error bounds are often difficult to guarantee in practice, and accurate observers for highly nonlinear systems are rarely feasible.
\begin{lemma}\label{lemma_1}
For the partially observed ct-SJDS $S$, if the solution to the optimization problem defined in \eqref{op_prob} is obtained, then the probability that the solution process $x_{x_0,v}$ of the system $S$ starting at $x_0$ reaches the unsafe region $X_u$ under the control policy $v$ within time horizon $[0,T)$ is upper bounded by $\frac{\gamma_1}{\gamma_2}+cT.$ 
\end{lemma}

However, due to the continuous state set $X$, there are infinitely many constraints in \eqref{op_prob}, and therefore it is computationally challenging to find a solution to the optimization problem. We overcome this by sampling a finite number of sample points from the augmented continuous state space set $X\times X$. We sample $N$ data points ($s^{(r)} = (x^{(r)}, \hat{x}^{(r)}), r\in[1;N]$) from the augmented state set. These data samples are generated by selecting a discretization parameter $\epsilon$, and sampling $N$ data points, which are the centers of the discretized grid cells, such that the following holds: %$\forall s\in X\times X, \exists s^{(r)}\in X\times X, r\in[1;N], \text{ s.t. } ||s-s^{(r)}||\leq \epsilon.$
\begin{align}
    \forall s\in X\times X, & \exists s^{(r)}\in X\times X, r\in[1;N], \text{ s.t. }\notag \\
    &||s-s^{(r)}||\leq \epsilon. \label{epsilon}
\end{align}

Additionally, since some states are unmeasurable and the state observer is itself unknown, which makes it difficult to fix the template of CBF, controller, and observer dynamics, we employ neural networks to approximate the functions for CBF, observer, and controller. We represent these neural network-based CBF (NCBF), controller, and observer as $B_{\theta_1}, \textsl{g}_{\theta_2}, \hat{f}_{\theta_3}$, where $\theta_1, \theta_2, \theta_3$ are the trainable parameters of the neural network. 
We denote the neural network architecture as $\{n^0, \{n^l\}^l,n_o\}$ with an input layer of $n^0$ neurons, $l$ number of hidden layers, each with $n^l$ neurons, and an output layer of $n_o$ neurons. The activation function applied to neurons is indicated by $\sigma(\cdot)$. Since we require the computation of derivatives and the Hessian matrix of CBF, a twice-differentiable activation function is preferred (for example, Softplus, Tanh, Sigmoid, etc.) for NCBF. One can refer to \cite{NN_observer_cts_time} for the detailed neural network architecture.

\subsection{Formulation of Loss Functions}
Now we frame the optimization constraints as suitable loss functions for training the NCBF, controller, and observer, which, when minimized, result in satisfaction of conditions in \eqref{op_prob}.
Consider the given partially observed system $S$ of the form \eqref{dyn}, an initial set $X_0\subset X$, and an unsafe set $X_u\subset X$. After sampling from the augmented state space $X\times X$, we form the following sets: $\mathcal{X}_d=\{s^{(r)}\mid s^{(r)}\in X\times X, r\in [1;N]\}, ${$\mathcal{X}_0=\mathcal{X}_d\cap (X_0\times X_0), \mathcal{X}_u=\mathcal{X}_d\cap (X_u\times X)$.} We consider $\mathcal{X}_0 \cap \mathcal{X}_u =\emptyset$.

All three neural networks $B_{\theta_1}, \textsl{g}_{\theta_2}$, and $\hat{f}_{\theta_3} $ are trained such that the following sub-loss functions are minimized:
\begin{align}
    &L_1(\theta_1) = \sum_{s^{(r)}\in \mathcal{X}_0} ReLU\big( B_{\theta_1}(s^{(r)}) - \gamma_1\big), \label{l1}\\
    &L_2(\theta_1) = \sum_{s^{(r)}\in \mathcal{X}_u} ReLU\big(-B_{\theta_1}(s^{(r)}) + \gamma_2\big), \label{l2}
     \\
    & L_3 (\theta_1, \theta_2, \theta_3) = \sum_{s^{(r)}\in \mathcal{X}_d} ReLU({\mathcal{D} B}(s^{(r)}, \textsl{g}_{\theta_2}(\hat{x}^{(r)})) - c)\notag \\
    & = \sum_{s^{(r)}\in \mathcal{X}_d} ReLU \bigg(\partial_x B_{\theta_1}f(x^{(r)},\textsl{g}_{\theta_2}(\hat x^{(r)}))+\partial_{\hat x^{(r)}}B_{\theta_1}
    \hat{f}_{\theta_3}\big(\hat{x}^{(r)}\hspace{-0.05cm},\textsl{g}_{\theta_2}(\hat x^{(r)}), h(x)-h(\hat x)\big)+\hspace{-0.05cm}\frac{1}{2}\text{Tr}(g_1g_1^\top\partial_{xx}B_{\theta_1})\nonumber
    \\
    &\quad+\sum_{i=1}^{\overline{q}_1}\lambda_{1i}(B_{\theta_1}(x^{(r)}+r_1(x^{(r)})\mathbf{e}_i, \hat x^{(r)})-B_{\theta_1}(s^{(r)}))-c\bigg),\label{l3}
\end{align}
where $ReLU(z) =\max (0,z)$.

In addition to these sub-loss functions, we also define a loss function for the observer neural network based on the Lyapunov function defined over the estimation error, in order not to produce random values for the state estimation. 

Let us denote the estimation error as $e = x-\hat{x}$. The error dynamics is $\text{d}e = (f(x,\textsl{g}_{\theta_2}(\hat{x}))-\hat{f}_{\theta_3}(\hat{x},\textsl{g}_{\theta_2}(\hat{x}),h(x)-h(\hat x)))\text{d}t + g_1(x)\text{d}W_{1t}+r_1(x)\text{d}P_{1t}$.

Now, let a candidate continuous Lyapunov function $V:\mathbb{R}^n\rightarrow \mathbb{R}_{\geq 0}$ be defined as $V(e)=\frac{1}{2}e^\top e $, with the following properties \cite{wang2017observer}:
     $\forall x,\hat{x}\in \mathcal{X}_d$, $\mathcal{D}{V} = e^\top \big(f(x,\textsl{g}_{\theta_2}(\hat{x})) -\hat{f}_{\theta_3}(\hat{x},\textsl{g}_{\theta_2}(\hat{x}), h(x)-h(\hat x))\big) +\frac{1}{2}\text{Tr}(g_1g_1^\top) + \sum_{i=1}^{\overline{q}_1}\lambda_{1i}(V(e+r_1(x)\mathbf{e}_i)-V(e)) 
     %+ \sum_{i=1}^{\overline{q}_2}\lambda_{2i}(V(e+r_2\mathbf{e}_i)-V(e))
     \leq -\alpha V(e)$, where $\alpha>0$.
One can reformulate the above property as the following loss function for the observer neural network.
\begin{align}
    L_{4}(\theta_3) =& \sum_{s^{(r)}\in\mathcal{X}_d}ReLU\bigg(e^\top \big(f(x^{(r)},\textsl{g}_{\theta_2}(\hat{x}^{(r)}))-\hat{f}_{\theta_3}\big(\hat{x}^{(r)},\textsl{g}_{\theta_2}(\hat{x}^{(r)}), h(x^{(r)})-h(\hat{x}^{(r)})\big)\big) +\frac{1}{2}\text{Tr}(g_1g_1^\top) \nonumber\\&+ \sum_{i=1}^{\overline{q}_1}\lambda_{1i}(V(e+r_1(x^{(r)}) \mathbf{e}_i)-V(e))
    %\nonumber\\&+\sum_{i=1}^{\overline{q}_2}\lambda_{2i}(V(e+r_2 \mathbf{e}_i)-V(e))
    +\alpha V(e)\bigg).\label{l_obs}
\end{align}
The overall loss function to be minimized while training the neural network is the weighted sum of the above sub-loss functions and is computed as follows:
\begin{align}
    L_{cbf}(\Theta) &= k_1L_1(\Theta)+k_2L_2(\Theta)+k_3L_3(\Theta),%+k_4L_4 
    \label{cbf_loss_fn_nn}\\
    L_{obs} &= k_{4}L_{4}(\theta_3), \label{obs_loss_fn}
\end{align}
where $k_1,k_2,k_3, k_4>0$ are the positive real-valued weights corresponding to the sub-loss functions, and $\Theta=[\theta_1, \theta_2, \theta_3]$ are the trainable parameters of the neural networks. We propose the optimization objective as another loss function:
\begin{align}
    L_5(\eta) = ReLU(c-\beta(\gamma_2-\gamma_1)),
\end{align}
where $\eta=[\gamma_1,\gamma_2, c]$ are the parameters of CBF conditions in \eqref{op_prob}. Now we consider the following assumption on Lipschitz continuity, which will help to provide a formal guarantee as discussed in the next subsection.
\begin{assumption}
    The candidate NCBF is assumed to be Lipschitz continuous with Lipschitz bound $L_B$. Its derivative is represented by a slightly modified neural network with changes reflected in the weights of the last layer. It is Lipschitz bounded by a constant $L_{\partial B}$ [refer to \cite{Basu02022026} for more details]. The controller and observer neural networks have Lipschitz bounds $L_c$ and $L_o$, respectively. Additionally, the partial derivative $\frac{\partial B}{\partial x}$, functions $f(x,u)$, $\hat{f}_{\theta_3}$ and $h(x)$ are bounded by $M_{\partial B}, M_f, M_o, M_h$ respectively, i.e., $\sup_x \lVert \frac{\partial B}{\partial x} \rVert \leq M_{\partial B}, \sup_{(x,u)} \lVert f(x,u)\rVert \leq M_f$,  $\sup_{(\hat{x},u)} \lVert \hat{f}_{\theta_3}(\hat{x},u,y-\hat{y})\rVert \leq M_o$, $\Vert\sup_x h(x)\Vert\leq M_h$. \label{assumption1}
\end{assumption}

Assumption \ref{assumption1} over the Lipschitz constants of the trained neural network $B_{\theta_1}$ and its derivative $\partial_x B$ and Hessian $\partial^2B$ are ensured by using the Lemma adopted from \cite[Theorem 2]{fazlyab2019efficient}. To satisfy the LMIs corresponding to the Lipschitz constant of the neural networks of CBF, observer and controller, we formulate the loss function:
\begin{align}
    &L_M(\Theta, \Lambda) = -c_{l_1}\log\det(M_{L_B}(\theta_1, \Lambda_1)) \nonumber\\&-c_{l_2}\log\det(M_{L_{\partial B}}(\hat{\theta}_1, \hat\Lambda_1)) - c_{l_3}\log\det(M_{L_{\partial^2 B}}(\tilde\theta_1, \tilde\Lambda_1))\nonumber\\
    & \!-\! c_{l_4}\log\det(\!M_{L_c}\!(\theta_2, \Lambda_2)) \!-\! c_{l_5}\log\det(\!M_{L_o}\!(\theta_3, \Lambda_3)), \label{lmi_loss}
\end{align}
where $c_{l_1},c_{l_2},c_{l_3}, c_{l_4}, c_{l_5}$ are positive real-valued weights for sub-loss LMIs, $\Lambda = [\Lambda_1, \hat\Lambda_1, \tilde\Lambda_1, \Lambda_2,\Lambda_3]$ and $M_{L_B}(\theta_1,\Lambda_1), M_{L_{\partial B}}(\hat \theta_1,\hat \Lambda_1), M_{L_{\partial^2B}}(\tilde\theta_1,\tilde\Lambda_1), M_{L_c}(\theta_2,\Lambda_2)$, $M_{L_o}(\theta_3,\Lambda_3)$ are the matrices corresponding to the bounds $L_B, L_{\partial B}, L_{\partial^2B}, L_c, L_o$, respectively.

\subsection{Training of Neural Networks with Formal Guarantees}

Although the optimization problem is approximated using a finite set of sampled points, it is necessary to establish that the resulting solution with the sampled points preserves the probabilistic safety guarantee of the original optimization problem. To this extent, we provide the following lemma.

\begin{lemma} \cite[Ex 3.3]{khalil2002nonlinear} \label{lemma_khalil}
    If two functions $f_1$ and $f_2$ are Lipschitz continuous with constants $L_{f_1}$ and $L_{f_2}$, respectively, and are bounded by $\sup \lVert f_1\rVert \leq M_{f_1}$ and $\sup \lVert f_2 \rVert \leq M_{f_2}$, then their product $f_1f_2$ is also Lipschitz continuous with Lipschitz constant $M_{f_1}L_{f_2}+M_{f_2}L_{f_1}$. 
\end{lemma}

Now we propose the theorem that establishes the probability of the system trajectory evolving under the controller trained with a finite number of sampled points from the augmented state space.

\begin{theorem}\label{th:validity}
    Consider a partially observed ct-SJDS \eqref{dyn} with state space $X$, an initial state set $X_0\subseteq X$, and an unsafe set $X_u\subseteq X$. Let $B_{\theta_1}(x,\hat{x})$ be the neural network-based CBF under the controller $\textsl{g}_{\theta_2}$ and observer $\hat{f}_{\theta_3}$ with trainable parameters $\theta_1, \theta_2, \theta_3$. According to Assumption~\ref{assumption1}, the functions $q_k(x, \hat{x}), k\in\{1,2,3\}$ in \eqref{op_prob} are Lipschitz continuous with $L_{q_1} = L_{q_2}=L_B, L_{q_3} =L_{\mathcal{D}B}=\max(L_x+L_uL_c, L_o(1+L_c+L_h)) + L_{\partial B}\sqrt{M_f^2+M_o^2}+M_{\partial B}L_F + 0.5(\lVert g_1g_1^\top\rVert L_{\partial_{xx}B}+\lVert\partial_{xx}B\rVert L_{g_1g_1^\top}) +2\sqrt{2}L_B \sum_i\lambda_{1i}(\lVert r_1\mathbf{e}_i\rVert)$ and satisfy the following condition. \begin{align}
       \hspace{-0.3cm} \max(q_i(s^{(r)})) \hspace{-0.1cm} \leq   \hspace{-0.1cm}  0, i \hspace{-0.1cm} \in \hspace{-0.1cm}[1;3],\forall s^{(r)}\in X\times X, \forall r\in[1;N]. \label{validity_condition}
    \end{align}
    Then the probability that the solution process $x_{x_0, \textsl{g}_{\theta_2}}$ of the system starting at an  initial state $x_0\in X_0 \ \mathbb{P}$-a.s. reaches the unsafe region $X_u$ under the controller $\textsl{g}_{\theta_2}$ within the time duration $[0,T)$ is upper bounded by $\frac{\gamma_1+L_B\epsilon}{\gamma_2-L_B\epsilon}+(c+L_{\mathcal{D}B}\epsilon)T$.%\textcolor{blue}{, where $\gamma_2>\gamma_1+2L_B\epsilon$}.
\end{theorem}

\begin{proof}
Using \eqref{epsilon}, Assumption \ref{assumption1}, Lemma \ref{lemma_1} and \ref{lemma_khalil}, we obtain:
\begin{align*}
    % &\forall s\in X_0\times X_0, \exists s^{(r)}\in X_0\times X_0, r\in [1;N],\\
    % &\text{s.t. }||s-s^{(r)}\leq\epsilon
   &\forall s\in X_0\times X_0, \exists s^{(r)}\in X_0\times X_0, \\&B(s)=B(s)-B(s^{(r)}) + B(s^{(r)})\nonumber\\& \quad\quad\leq L_{B}||s-s^{(r)}||+\gamma_1
   \leq L_{B}\epsilon +\gamma_1,
   \\
   &\forall s\in X_u\times X, \exists s^{(r)}\in X_u\times X, \\
   &-B(s) =-B(s)+B(s^{(r)})-B(s^{(r)})\leq L_B\epsilon -\gamma_2,
   % \\ 
   \end{align*}
    \begin{align*}
   &\forall s\in X\times X, \exists s^{(r)}\in X\times X,\\
   &\mathcal{D}B(s,u)=\mathcal{D}B(s,u)-\mathcal{D}B(s^{(r)},u)+\mathcal{D}B(s^{(r)},u)\\
   &\quad\quad\quad\quad\leq L_{\mathcal{D}B}\epsilon+c.
\label{proof_q_lipschitz}
\end{align*}
where $L_{q_1}=L_{q_2}=L_B, L_{q_3}=L_{\mathcal{D}B}=\max(L_x+L_uL_c, L_o(1+L_c+L_h)) + L_{\partial B}\sqrt{M_f^2+M_o^2}+M_{\partial B}L_F + 0.5(\lVert g_1g_1^\top\rVert L_{\partial_{xx}B}+\lVert\partial_{xx}B\rVert L_{g_1g_1^\top}) + 2\sqrt{2}L_B (\sum_i\lambda_{1i}\lVert r_1\mathbf{e}_i\rVert)$. The satisfaction of \eqref{validity_condition} implies the satisfaction of Lemma \ref{lemma_1} with $\gamma_1, \gamma_2, c$ replaced by $\gamma_1+L_B\epsilon$, $\gamma_2-L_B\epsilon$, $c+L_{\mathcal{D}B}\epsilon$, respectively. Now, substituting the new values of $\gamma_1, \gamma_2, c$ in Theorem \ref{th:barrier}, we obtain the upper bound on the probability $\mathbb{P}\{x_{x_0,u}(t)\in X_u, \exists t\in[0,T)\mid x_0, \hat{x}_0\}\leq \frac{\gamma_1+L_B\epsilon}{\gamma_2-L_B\epsilon}+(c+L_{\mathcal{D}B}\epsilon) T$.
\end{proof}

Now we propose the main algorithm to jointly synthesize the NCBF, the controller and the observer.
\begin{algorithm}
\caption{NN Training}
\label{NN_training}
\begin{algorithmic}[1]
    \Require System dynamics $S$, Sets $X_0, X_u, X$, Data sets: $s^{(r)}\in  X\times X, r\in [1;N]$
    %, Lipschitz bounds: 
    \Ensure $B_{\theta_1}, \textsl{g}_{\theta_2}, \hat{f}_{\theta_3}, \gamma_1, \gamma_2, c $
    \State Select Epochs, NN hyperparameters ($l, n^l$, activation function, optimizer, scheduler), $k_1, k_2, k_3, k_{4}$
    \State Initialize trainable parameters ($\Theta, \eta, \Lambda$)
    \For{$i\leq Epochs$ (Training starts here)}
        \State Create batches of training data from $\mathcal{X}_d, \mathcal{X}_0, \mathcal{X}_u$
        \State Find batch loss $L=L_{cbf}+L_{obs}+L_5+L_M$ using \eqref{l1}-\eqref{lmi_loss}.
        \State Use ADAM optimizer \cite{Adam} to update $\Theta^i, \eta^i, \Lambda^i$
        \State Find the Lipschitz constants of the networks using the ECLipsE tool \cite{eclipse}
    \EndFor
    \State \textbf{return} $B_{\theta_1}, \textsl{g}_{\theta_2}, \hat{f}_{\theta_3}, \gamma_1, \gamma_2,c $
\end{algorithmic}
\end{algorithm}

We now define the main theorem, whose satisfaction will guarantee that the probability of the system starting in $X_0$ under the trained controller $g_{\theta_2}$ reaching the unsafe set is upper-bounded as stated in Theorem \ref{th:validity}.

\begin{theorem}\label{th:convergence}
    Consider a partially observed ct-SJDS \eqref{dyn} with state space $X$, an initial set $X_0$, and an unsafe set $X_u$. Let $B_{\theta_1}(x,\hat{x})$ be the neural network CBF with a controller network $\textsl{g}_{\theta_2}(\hat{x})$, and an observer network $\hat{f}_{\theta_3}(\hat{x}, \textsl{g}(\hat x), h(x)-h(\hat{x}))$, trained over finite data points, such that the loss functions $L_{cbf}(\Theta)=0, L_M(\Theta, \Lambda)\leq 0$ and $L_{obs}$ is minimized. Then, starting at any initial state in $X_0$, we obtain an upper bound on the probability that the state trajectories enter the unsafe region in the time interval $[0,T)$, given by $\frac{\gamma_1+L_B\epsilon}{\gamma_2-L_B\epsilon}+(c+L_{\mathcal{D}B}\epsilon)T$ as per Theorem \ref{th:barrier}. Consequently, the probability of the state trajectories starting at some initial state $x_0\in X_0$ remaining safe outside the unsafe region is given as: $\mathbb{P}\{x_{x_0,\textsl{g}}(t)\notin X_u,\forall t\in[0,T)\mid x_0\}>1-\bigg( \frac{\gamma_1+L_B\epsilon}{\gamma_2-L_B\epsilon}+(c+L_{\mathcal{D}B}\epsilon)T\bigg)$.
\end{theorem}
\begin{proof}
The loss $L_{cbf}(\Theta)=0, L_M(\Theta, \Lambda)\leq 0$, implies that the conditions in Theorem \ref{th:validity} have been met with appropriate Lipschitz constants. Hence, through the trained controller $\textsl{g}_{\theta_2}$, we obtain the upper bound of the probability $\mathbb{P}\{x_{x_0,\textsl{g}}(t)\in X_u,\exists t\in[0,T)\mid x_0\}\leq \frac{\gamma_1+L_B\epsilon}{\gamma_2-L_B\epsilon}+(c+L_{\mathcal{D}B}\epsilon)T$ as discussed in Theorem \ref{th:barrier}. 
\end{proof}
{
\begin{remark}
    Since we trained NCBF $B_{\theta_1}$ on the data set of the augmented system, we do not require that the observer loss $L_{obs}$ in \eqref{obs_loss_fn} converge to zero for synthesizing the controller $\textsl{g}_{\theta_2}$. Even with a nonzero $L_{obs}$, Theorem \ref{th:convergence} holds. This can be readily observed in case studies in Section \ref{Sec:Simul}.
\end{remark}
}
% \begin{remark}
%     {The algorithm lacks a general convergence guarantee, but strategies like reducing the discretization parameter $\epsilon$ \cite{zhao2021learning} or adjusting neural network hyperparameters (architecture, learning rate) \cite{nn_lr} can improve convergence.}
% \end{remark}

\section{Simulation Results}\label{Sec:Simul}
In this section, we validate our proposed methodology through 2 simulations: the linear DC motor model and the nonlinear model. \\
% \textit{A. Jet Engine: } First, we consider an example of a nonlinear ct-SJDS of jet engine model with the following dyanmics:
% \begin{align*}
%     \text{d}{x_1}&= (-x_2-\frac{3}{2}x_1^2-\frac{1}{2}x_1^3)\text{d}t+0.02\text{d}W_{11t} + 0.09\text{d}P_t,\\
%     \text{d}x_2&= u\text{d}t+0.06\text{d}W_{12t},\\
%     \text{d}y&=x_2\text{d}t+0.06\text{d}W_{2t},
% \end{align*}
% where $x_1, x_2$ are the states and $x_2$ is the observed output. The input to the system is $u$. The terms $W_{11t}, W_{12t}, W_{2t}$ are the standard Brownian motions and $P_t$ is the Poisson process with rate $\lambda=0.1$. We consider a state space $X=[-0.5, 0.5]\times[-0.5, 0.5]$, and unsafe set $X_u= X\setminus([-0.4,0.4]\times[-0.4,0.4])$. 

\begin{figure}
    \centering
    \includegraphics[width=0.95\linewidth]{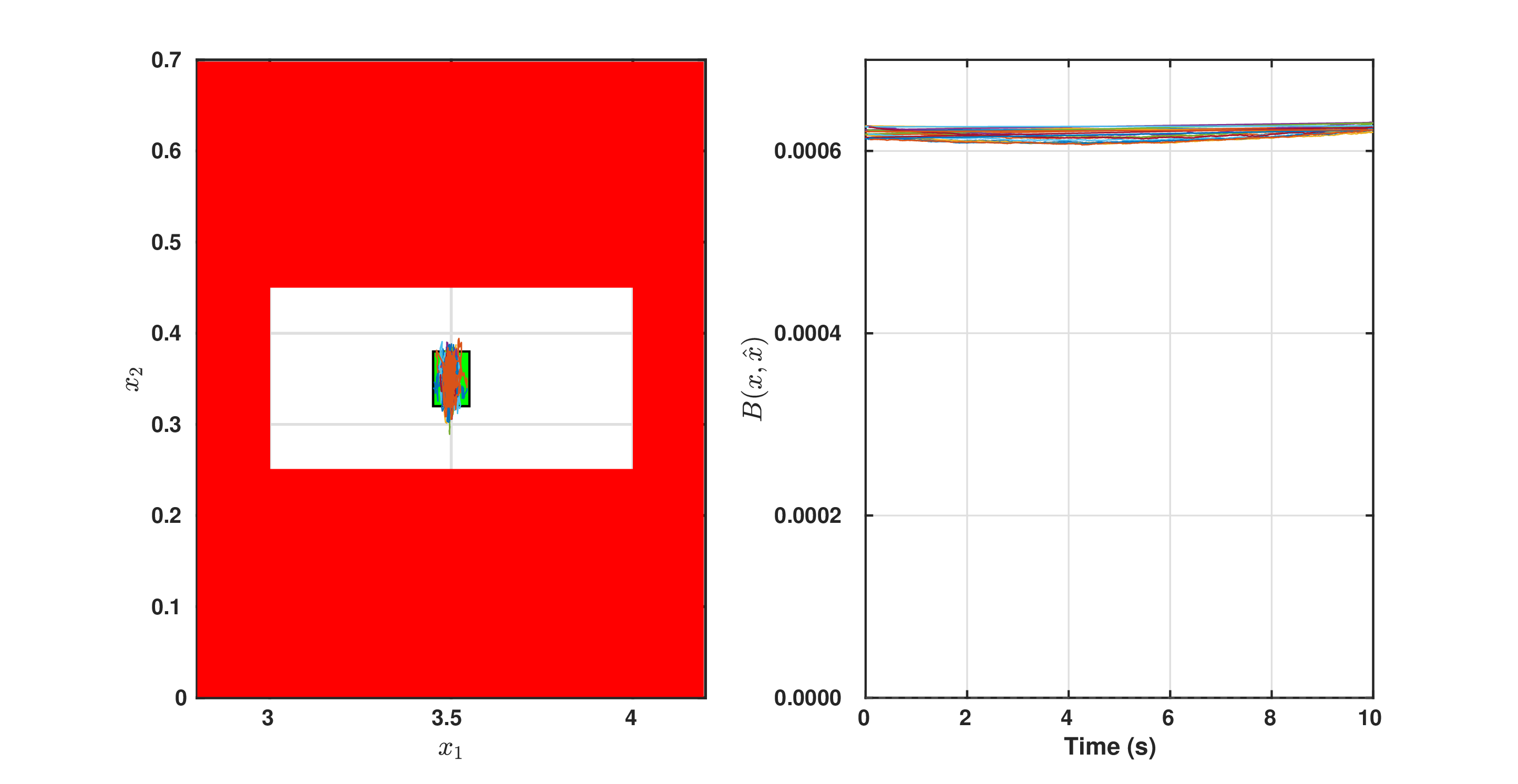}
    \caption{DC Motor (a) Actual trajectories starting from different initial conditions $x_0 \in X_0$ (green region) and away from the unsafe region (red region) (b) $B(x,\hat x)< \gamma_1=0.001$ for all $t\in[0,10]$ seconds.}
    \label{fig:dc_motor}
\end{figure}

{\textit{A.} \textit{DC Motor}:} First, we consider an example of a DC motor \cite{obeidat2013real} with the following dynamics {of the form \eqref{dyn}}:
\begin{align*}
\text{d}{x_1}&= (\frac{-R}{L}x_1 - \frac{K_{dc}}{L}x_2 + \frac{V_{in}}{L})\text{d}t+0.02\text{d}W_{11t} + 0.009\text{d}P_t,\\
    \text{d}x_2&= (\frac{K_{dc}}{J}x_1 - \frac{b}{J}x_2)\text{d}t+0.06\text{d}W_{12t},\\
    \text{d}y&=x_2\text{d}t,%+0.06\text{d}W_{2t},
\end{align*}
where $x_1$, $x_2$ are the armature current and rotational speed, and the observed output is the speed $x_2$. The input to the system is a voltage source $V_{in}$. The motor parameters considered are $J=0.01 kg.m^2$, $b=0.1 N.m.s$, $K_{dc}=0.01 V/rad/sec$, $R=1 \Omega$, $L=0.5H$.
We consider the state space $X = [2.8,4.2]\times[0.0, 0.7]$, an initial set $X_0=[3.45,3.55]\times[0.32,0.38]$, and an unsafe set $X_u=X\setminus[3.0,4.0]\times[0.15,0.55]$. The discretization is set as $\epsilon=0.017$. We fix the three neural network architectures with $n^0, l, n^l, n_o$ as the number of input neurons, hidden layers, neurons in each hidden layer, and output neurons, respectively. NCBF: $n^0=4, l=2, n^l=\{64,64\}, n_o=1$. NN observer: $n^0=4, l=1, n^l=\{8\}, n_o=2$, NN controller: $n^0=2, l=1, n^l=\{64\}, n_o=1$. The activation function used in hidden layers is Tanh.
The training algorithm converges to obtain neural networks for the observer $(\hat{f}_{\theta_3})$, NCBF $(B_{\theta_1})$, and the controller $(\textsl{g}_{\theta_2})$ with $\gamma_1=0.001, \gamma_2=1.1001,c=0.0001$. 
The Lipschitz constants of the neural networks computed using the ECLipsE tool \cite{eclipse} are $L_{B}=0.00525, L_{o}=0.0148, L_c=0.006$. Figure \ref{fig:dc_motor} shows 10 actual trajectories for $T=10$ seconds, starting with different initial conditions. As per Theorem \ref{th:validity}, we computed the probabilities that trajectories starting anywhere in $X_0$ reach the unsafe set in time $T=0.5,1,2$ seconds are upper bounded by $0.215, 0.43, 0.858$, respectively.
From Figure \ref{fig:dc_motor}, we can observe that the trajectories remain inside the safe set and away from the unsafe region (red color region) while maintaining barrier value $B(x,\hat x)<\gamma_1$.
\\

\begin{figure}
    \centering
    \includegraphics[trim=4cm 0cm 2cm 1cm, clip, width=0.9\linewidth]{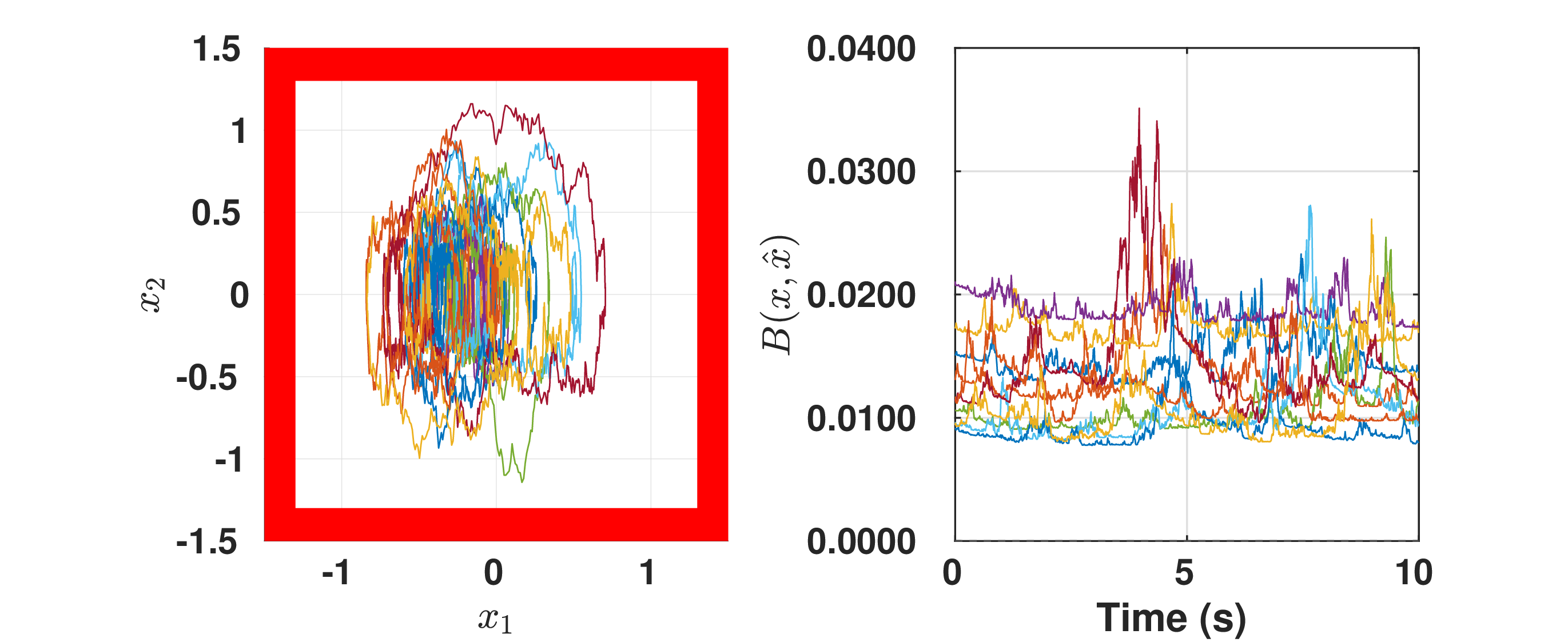}
    \caption{Nonlinear system: Actual state trajectories starting in $X_0$ and the barrier value $B(x, \hat x)\leq \gamma_1=0.1$.}
    \label{fig:nonlinear_sys}
\end{figure}

\textit{B. Nonlinear System}:
We consider a simulation example:
$\text{d}x_1 = x_2\text{d}t, \text{d}x_2 = (-x_1-x_2-x_1^3+u)\text{d}t+0.5\text{d}W_t.$
We consider $\text{d}y=x_2\text{d}t$. Let the state space $X=[-1.5,1.5]\times[-1.5,1.5]$, initial region $X_0=\{(x_1,x_2)\mid (x_1+0.5)^2+x_2^2\leq 0.12^2\}$, unsafe region $X_u=X\setminus [-1.3,1.3]\times[-1.3,1.3]$. The discretization is set as $\epsilon=0.01$. NCBF: $n^0=4, l=2, n^l=\{20,20\}, n_o=1$. NN observer: $n^0=4, l=1, n^l=\{20\}, n_o=2$, NN controller: $n^0=2, l=2, n^l=\{20,20\}, n_o=1$. We selected Softplus as the activation function for the hidden layers. The training algorithm converges to obtain neural networks for the observer $(\hat{f}_{\theta_3})$, NCBF $(B_{\theta_1})$, and the controller $(\textsl{g}_{\theta_2})$ with $\gamma_1=0.1, \gamma_2=0.5,c=0.01$. The Lipschitz constants calculated using the ECLipsE tool are $L_B=0.021, L_o=10^{-5}, L_c=0.54$. The value of $L_{\mathcal{D}B}$ as calculated by the formula in Theorem \ref{th:convergence} is 3.044, and the probability of state trajectories starting in $X_0$, entering the unsafe set under the trained neural network controller policy within 0.5, 1, 5, 10 seconds are upper bounded by $0.221, 0.241,0.403, 0.605$, respectively. Figure \ref{fig:nonlinear_sys} shows the state trajectories for 10 runs satisfying $B(x,\hat x)\leq \gamma_1$.

\section{Conclusion}
This paper presented a framework for the co-synthesis of a neural network control barrier function and an output-feedback-based controller for partially observed continuous-time stochastic systems, providing probabilistic safety guarantees. A sampled optimization framework using a neural network approach was developed to ensure the probability of state trajectories starting in an initial region, reaching an unsafe region is upper-bounded. Simulation results demonstrated the effectiveness of the proposed approach in maintaining the safety of stochastic systems.

\bibliographystyle{IEEEtran}
\bibliography{reference.bib}
\end{document}